\documentclass[twocolumn]{article}

\usepackage{cite}
\usepackage{amsmath}
\usepackage{amsfonts}
\usepackage{algorithmic}
\usepackage{array}
\usepackage{hyperref}
\usepackage{cancel}

\usepackage{fixltx2e}
\usepackage{graphicx}
\usepackage{url}
\usepackage{xcolor}
\usepackage{float}

\newcommand{\myhref}[2]{\textcolor{blue}{\href{#1}{#2}}}
\newcommand{\myurl}[1]{\textcolor{blue}{\url{#1}}}

\newcommand{\ignore}[1]{}

\begin{document}
\title{Applications of Recursively Defined Data
  Structures\footnote{
    Copyright\copyright 1993,
    Australian Computer Society Inc. General permission
    to republish, but not for profit, all or part of this material is granted,
    provided that the ACJ’s copyright notice is given and that reference is
    made to the publication, to its date of issue, and to the fact that
    reprinting privileges were granted by permission of
    the Australian Computer Society Inc.
    Manuscript received: November 1991, revised February and July 1992.\newline
    (Originally appeared in the
    Australian Computer Journal, vol.25, no.1, pp.14-20,
    February 1993.)
  }
}
\author{ Lloyd Allison, \\
  Department of Computer Science, \\
  Monash University, Clayton, Victoria 3800, Australia \\
  lloyd.allison@monash.edu \\
}

\date{}

\maketitle

\begin{abstract}
A circular program contains a data structure whose
definition is self-referential or recursive. The use of such a
definition allows efficient functional programs to be written
and can avoid repeated evaluations and the creation of
intermediate data structures that would have to be garbage
collected. This paper uses circular programs in various
ways, to implement memo-structures and explicit search-trees
to hold solutions to constraint-satisfaction problems.
\end{abstract}

\section{Introduction}
\label{sec:Intro}

A circular program contains a data structure whose definition is
self-referential or recursive. Such a program cannot be written
in a conventional, strict, imperative programming language but
it can be written in a functional language employing lazy
evaluation \cite{FW76,Hen76} or call by need.

\begin{figure}[H]
\begin{verbatim}
  general schema: let rec ds = f(ds)
     -- ds is some data structure

  eg let rec posints = 1.(map succ posints)
     -- list of all +ve integers
\end{verbatim}
\end{figure}

Note that `.' is the infix list constructor also known as cons,
rec qualifies recursive definitions and map applies a function to
each element of a list and so produces a new list.

The list posints contains all the positive integers
1.(2.(3...)) or
[1,2,3,...].
It begins with 1 and continues with the result of
applying the successor function succ to
each element of posints itself.
Successor applied to the first element gives the second
element, 2, and so on.
It is the definition of the \textit{value} of the data
structure posints, not just its type (list), being recursive that
makes this a circular program.

Under lazy evaluation, an expression is not evaluated unless
it is needed. In particular, the right hand side of a definition, and
the actual parameter of a function, are not evaluated until they
are needed -- if they are needed. When an expression is
evaluated, the value is remembered to avoid recomputation later.
(The conditional, `if', is the only non-strict or lazy operator
in many imperative languages.) Lazy evaluation permits recursive
definitions of data structures and also allows some computations
with infinite data structures.
\textit{All} of a potentially infinite
data structure can be defined although only a finite part may be
evaluated. If only a bounded part were defined and evaluated, a
copy would have to be made if it had to be extended, wasting
time and space, because functional languages do not permit side-effects.
Circular programs can, in certain cases, have it both
ways -- an expanding data structure with side-effect-free programming.
As a bonus, infinite data structures are sometimes
easier to define because boundary cases are simpler or absent.
Although posints represents an infinite list, a program using
posints does not loop unless an attempt is made to print or
otherwise evaluate all of the list. The program terminates
provided that only finite parts of such structures are manipulated.

It is sometimes necessary to distinguish between the data
structure as seen by the programmer and as implemented by the
language system.
If the definition simply incorporates the data
structure directly, as in ones below, then a cyclic structure of
cells and pointers is created in the computer memory.

\begin{verbatim}
  let rec ones = 1 . ones  -- = [1,1,1,...]
\end{verbatim}

\noindent data structure in memory:

\begin{verbatim}
  ones: -----> 1.----->|
               ^       |
               |       v
               |<-------
\end{verbatim}

This ability to create cyclic structures can be used to form
circular lists, doubly linked lists and threaded trees in a
functional language \cite{All89}.
The programmer cannot
determine if a cyclic structure has been formed, except indirectly
by the program’s speed or modest use of space.
If the
recursive definition uses some function of the contents of the
data structure, as in posints, no cyclic structure is created at the
implementation level but those parts already computed can be
used to compute new parts. This can be used to implement
queues \cite{All89} and various space efficient programs.

Many functional programs compute their final result in
stages, some data structure being operated on, often in small
steps or passes, by functions such as map, filter, reduce, and
so on.
Each step produces an intermediate data structure
which is eventually discarded and collected as garbage at
some cost.
Bird \cite{Bir84} used circular programs in program
transformations to convert multi-pass algorithms into
single-pass algorithms.
He attributed knowledge of circular
programs to Hughes and to Wadler and the technique is so
useful that it has probably been discovered several times.

The objective of the paper is to promote this useful functional
programming technique.
The examples given here construct lists and trees in new ways.
They are used to define
memo-structures and explicit search-trees which remove the
need to repeat tests in certain constraint-satisfaction problems.
They have all been run on a small lazy
\myhref{https://www.cantab.net/users/mmlist/ll/FP/Lambda/}{interpreter}
which was instrumented to record program behaviour and they
perform as predicted. The notation used is from a hypothetical,
``generic'' functional language and is explained when required.

\section{Circular Lists}
\label{sec:CLists}

The ones and posints examples are amongst the simplest
circular programs. To introduce the technique more fully,
some non-trivial but routine examples on lists are given here.
A popular example computes the Hamming numbers.
These are all numbers of the form
$2^i \times 3^j \times 5^k, ~ i, j, k \ge 0$,
i.e.,
$1, 2, 3, 4, 5, 6, 8, 9, 10, 12, 15, ...$~.
In this section various circular programs
are derived from the Hamming numbers program.

The simplest program for the Hamming numbers is the following:

\begin{verbatim}
  let rec Hamming =
     1 . (merge3 (map (* 2) Hamming)
                 (map (* 3) Hamming)
                 (map (* 5) Hamming))
\end{verbatim}

This is an example used in many books and
papers \cite{BW88,Hen80,Tur84}.
Hamming is a self-referential list.
It begins with 1 followed by the result of merging three lists.
These are the results of multiplying all
members of Hamming by 2, 3 and 5 respectively
giving [2,...], [3,...] and [5,...].
The second Hamming number is thus 2, so the
three lists are [2,4,...], [3,...] and [5,...],
which allows the computation to proceed to the next step.
Many duplicate
numbers are produced, for example $6 = 2 \times 3 = 3 \times 2$,
and all but one copy must be removed by the merge function.
It is well known that this inefficiency can be
removed by ensuring that the factors in a product
are combined in ascending order.

\begin{verbatim}
 let rec
     a = 1. (map (*2) a)  -- [1,2,4,8,...]
     b = 1. (merge (tl a) (map (*3) b))
                 -- [1,2,3,4,6,8,9,12,...]
 and Hamming = 1.(merge(tl b)
                       (map (*5) Hamming))
\end{verbatim}

Note that hd (head) returns the first element of a list and tl
(tail) returns a list minus its first element.
A further operator null can be used to test if a list is empty.

The list a holds all powers of 2 and is defined in a very
similar way to posints.
List b holds all products of 2 and 3.
No duplicates are produced because powers of 2 are multiplied by
powers of 3 and then by powers of 5, in order.
Consequently a simpler version of merge can be used that
does not need to deal with duplicates.

It is a natural exercise to generalise the Hamming problem
to find all products of an arbitrary list of factors.
The factors are assumed to be coprime and in ascending order.

\begin{verbatim}
let rec
  products [] = [1] ||        --no factors
  products (f.fs) =  --at least 1 factor f
    let rec m=1.(merge (map (*f) m)
                       (tl (products fs)))
    in m
\end{verbatim}

Note this program uses pattern matching.
The function products accepts a single list parameter.
It distinguishes two cases, patterns or kinds of
input parameter:
the empty list [\,] and the general list f.fs
consisting of a first factor f and a list
of remaining factors fs.
A definition is given for each case.
Multiple cases are separated by $||$ which can be read as `otherwise'.

If there are no factors the list [1] is returned.
If there are
factors, the products of the first factor f and of all other factors
in fs must be combined.
The result list m is self-referential.
First, multiplying any member of m by f is itself a product --
map~(*f)~m.
Second, products of members of fs are also
members of m although the leading 1 is not needed --
tl~(products~fs).
Merging these two lists and putting 1 on the
front gives m.
It is easy to see that any valid product must be
produced by this process.

The last program works correctly on finite lists of factors.
It will not work when given an infinite list of factors because
the merge operation requires the head of two lists before it can
produce any output. For an infinite list of factors this would
require the heads of an infinite number of lists to be assembled
which is impossible.
This drawback can be overcome by
recognising that the second value in m must be the smallest
factor f itself.

\begin{verbatim}
 let rec
   products (f.fs) =    -- NB. not (null fs)
     let rec m=1.f.(merge(map (*f) (tl m))
                         (tl (products fs)))
   in m
\end{verbatim}

This program happens to require an infinite list of factors
although a case to allow for finite lists can be added.
Every
instance of m now has two values at the front before any merge
and thus one value remains when the tail is taken so output can
begin immediately.

\section{Circular Trees}
\label{Ctrees}

The previous section gave examples of circular programs over lists.
Here trees are defined in a similar way and used as
memo-structures to store the results of functions so that later
calls can access them quickly without recomputation.
In general, one or more functions and structures are defined
using mutual recursion.

\begin{verbatim}
  general schema:
    let rec         --mutually recursive
        ds = g ds   --data structure(s) and
    and f x = h x ds f --function(s)
\end{verbatim}

The technique is illustrated by application to the Fibonacci numbers.
Both the so called slow and fast Fibonacci programs
are well known. The slow version is doubly recursive and runs
in time exponential in~n:

\begin{verbatim}
  let rec slowfib n =
    if n<=2 then 1
            else slowfib(n-2)+slowfib(n-l)
\end{verbatim}

It is easily seen that the running time, $T(n)$, satisfies
$T(n)>2*T(n-2)$.
For example, slowfib(7) calls slowfib(5) and slowfib(6).
Slowfib(6) calls slowfib(5) and all its subcomputations again.
Many computations are repeated.
The fast program recognises that partial results are recalculated
many times by the slow program.
It gains efficiency by replacing binary recursion with
linear recursion to run in $O(n)$ time:

\begin{verbatim}
  let fastfib n =
    let rec f n a b =
      if n=1 then b
      else f(n-1) b (a+b)
    in f n 0 1
\end{verbatim}

The parameters a and b hold two successive Fibonacci
numbers. At the next step these become b and a+b respectively.

It is possible to define a circular program that builds a list
of the Fibonacci numbers:

\begin{verbatim}
  let rec
      fiblist = 1.1.(f fiblist)
            -- [1,1,2,3,5,...]
  and f(a.t) = (a+(hd t)).(f t)
\end{verbatim}

The function f both produces fiblist and uses it via its
parameter which always lags two steps behind the element
being calculated -- just enough.
It is now possible to find
fib(n) by indexing to the n\textsuperscript{th}
element of fiblist with the standard
function index:

\begin{verbatim}
let fib =
  let rec
     fiblist = 1.1.(f fiblist)  --memo list
  and f(a.t) = (a+(hd t)).(f t)
        --f builds fiblist
  and find n = index n fiblist
        --get nth element of fiblist
  in find
\end{verbatim}

The list fiblist is a structure storing old results of fib~n.
On a first call of fib~n the first n elements
of fiblist are constructed.
On a second call the result is just looked up in fiblist;
the second call is faster but is still $O(n)$ as index takes $O(n)$ time.
Actually, the result is looked up on the first call too, but it does
not yet exist and so the list is built to the required length.
Bird \cite{Bir80} discusses the use of arrays to store past values in the
process of deriving the fast Fibonacci program in an imperative language.
Hughes \cite{Hug85} describes a system that automatically stores
past values of functions for fast recall;
his system is implicit whereas the memo-structure here is explicit.

If a tree of Fibonacci numbers were built, the results of later
calls could be looked up in $O(log~n)$ time.
The nodes of a
complete binary tree can be numbered so that the children of
node n are 2n and 2n+1:

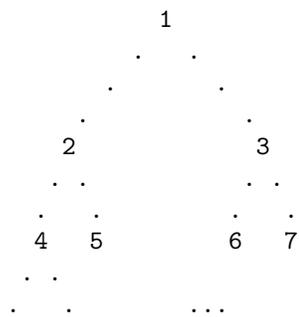
\begin{verbatim}
                 1
               .   .
             .       .
           .           .
          2             3
         . .           . .
        .   .         .   .
        4   5         6   7
       . .
      .   .        ...
\end{verbatim}

The value of fib n can be stored at node number n:

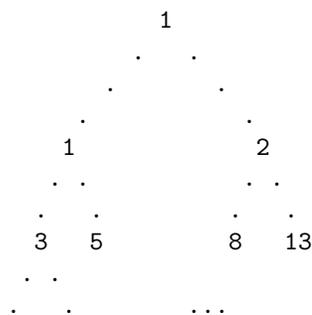
\begin{verbatim}
                 1
               .   .
             .       .
           .           .
          1             2
         . .           . .
        .   .         .   .
        3   5         8   13
       . .
      .   .        ...
\end{verbatim}

Given this tree and an integer, for example $n = 6_{10} = 110_2$,
the binary digits of n indicate whether to take left or right
subtrees in locating the n\textsuperscript{th}
node and this can be done in $O(log~n)$
time.
The bits of n are read from the \textit{second} most significant
bit to the least significant bit;
a 1 indicates go right and a 0 indicates go left.
Therefore $110_2$ implies: start at the root, go
right and then left.
We first define an infinite binary tree type:

\begin{verbatim}
  datatype tree = fork int tree tree
           --an infinite binary tree type

  let element (fork e l r) = e
        --extract the element value
  and left  (fork e l r) = l
        --extract left subtree
  and right (fork e l r) = r
        --extract right subtree
\end{verbatim}

Fork is a constructor that builds a new tree given an
element and left and right subtrees.
Element, left and right
return the components of a tree.

\begin{verbatim}
let fib =
  let rec
    fibtree = fork 1 (fork 1 (build 4)
                             (build 5))
                   (build 3) --fibtree:tree
              -- memo tree,
  and build n
        = fork (f(n-2)+f(n-l)) --build &f
               (build(2*n)) (build(2*n+1))
              -- construct fibtree
  and f n = lookup n element --return fib n
  and lookup 1 g = g fibtree || --decodes n
      lookup n g =              -- n>1
        lookup (n div 2)
               (g o(if even n then left
                              else right))
  in f
\end{verbatim}

Graphically, fibtree is defined to be:

\begin{verbatim}
                  1
                .   .
              .       .
            .          build 3
          1
         . .
        .   .
       .     build 5
      build 4
\end{verbatim}

The values in nodes one and two are both 1 and are
provided to enable node three to be built.
This allows node four to be built and so on.
The function lookup extracts the n\textsuperscript{th}
number from the tree.
It does this by constructing a function
g to follow left or right links according to the bits in n as described.
Functional composition `o' is used to link the
desired sequence of element, left and right operations together
and these are finally applied to fibtree.
$((p~o~q)(x) = p(q(x))).$

Assuming fib has been previously called with a parameter
greater than n, a second call fib~n takes $O(log~n)$ time to scan
down fibtree.
It might appear that the first such call would take
exponential time because of the two calls to f within build but
this is not the case.
The call f(n-2) causes the tree to be
evaluated and built up to node n-2.
The call f(n-1) only causes
one additional node to be evaluated using f(n-3) and f(n-2)
which are just looked up, their corresponding nodes having
been built already.
On the first call, there are $O(n)$ calls on
build, f and lookup the latter being logarithmic, the total time
taken is $O(n~log~n)$.
There is thus an increase in cost from $O(n)$
to $O(n~log~n)$ on first calls but a reduction from $O(n)$ to $O(log~n)$
on subsequent calls over the ``fast'' Fibonacci program.
There is also the cost of space to store the tree to be considered.

A subtle point should be noted: Our program
let fib = let rec ... in~f
binds fib to f which has fibtree in its environment so
fibtree persists for as long as fib does and is not recomputed.
If we carelessly defined
let fib2 n = let rec... in f~n
then fibtree
would persist for only as long as a call to fib2 remained
unevaluated and would be recomputed on each call.
Some
optimising compilers would undo this unfortunate effect by
effectively converting fib2 into fib.
However this is a difficult
issue because a programmer might deliberately write a function
having the form of fib2 because he or she needs a
temporary data structure but wants it to be destroyed to avoid
tying up space.

It is natural to ask if the log(n) factor in the costs can be
removed but it is caused by the use of a tree rather than an
(unbounded) array with $O(1)$ indexing.
In an imperative language, and in
some functional languages, one might use an
array instead of the tree.
However this would place a limit on the size of n.

If it is necessary both to have random access to the
Fibonacci numbers in $log(n)$ time and to have sequential
access then it may be convenient to derive a list from the tree
in breadth-first order. Since the tree is complete and infinite
the list is particularly simple to create:

\begin{verbatim}
let rec
    fibtree = as before
and build   = ...
and f       = ...
and lookup  = ...
and fiblist = map element nodes
  --elements in breadth-first order
and nodes
= bfirst fibtree
  --(sub)trees in breadth-first order
and bfirst t
  --perform breadth-first traversal of t
  let rec
      q = t.(traverse q)
  and traverse ((fork e l r).q2)
        = l.r.(traverse q2)
  in q
\end{verbatim}

Function bfirst returns a list or queue q of the subtrees of a
tree t in breadth-first order.
The queue begins with t itself and
the auxiliary function traverse produces the rest of the queue.
Function traverse examines the first element in q and adds its
subtrees to the end of q, i.e. in the second and third positions.
This is repeated for successive elements of q.
The definition of the queue q is self-referential.
Since q naturally grows as the
tree is scanned, and since it is also infinite, it is hard to see how
q could be created efficiently without a circular program.

The lists nodes and fiblist respectively contain the subtrees
and the elements of fibtree in breadth-first order.
Accessing the n\textsuperscript{th} element of fiblist also causes
fibtree to be evaluated to node number n.

\section{Circular Search Trees}
\label{CStrees}

Many search problems or \textit{constraint-satisfaction problems}
require finding a sequence of values
$\langle$a, b, c,...$\rangle$, or just abc...\,,
that satisfies certain \textit{constraints}.
A search program
explores the search space, building an implicit or explicit tree
of (partial) solutions.
If the constraints are uniform in a certain
sense then the solution tree may be defined recursively and
explicitly.
The advantage is that no test is performed twice as
the results of previous tests are available from the structure of
the tree.
This is valuable if the cost of performing tests
outweighs the cost of building and keeping the tree.

The uniformity required covers two conditions.
Firstly, it
must be possible to build all long solutions by extending short
solutions.
Secondly, the constraints on an element in the
sequence must involve other members of the sequence only in
ways that depend on their relative, not absolute, positions in
the sequence.
These conditions hold in many problems.

A suitable n-ary search-tree, in which a node contains an
element of type 't, a subtree and the siblings of the node, can
be defined as follows:

\begin{verbatim}
datatype tree 't = empty ||
                   node 't (tree 't)
                           (tree 't)
\end{verbatim}

Note that 't is a type parameter -- an arbitrary type.
There are two cases to tree -- the empty tree and a node -- separated
by $||$.
Siblings are linked together via the third component of a node.

As an example, the complete infinite tree over \{1,2,3\}
can be defined as:

\begin{verbatim}
let rec three =
  node 1 three
        (node 2 three
               (node 3 three empty))
  -- :tree int
\end{verbatim}

A simple generalisation of this example allows trees to be
built over the range [l..n]:

\begin{figure}[H]
\begin{verbatim}
  let build n =
    let rec
      T = toplevel 1 -- :tree int

    and toplevel m =
        if m > n then empty
        else node m (f T) (toplevel(m+1))

    and f T = T
    in T
\end{verbatim}
\end{figure}

The function toplevel builds the nodes at the top level of the
tree and f fills in the subtrees.
Here f is just the identity function
and as such it is redundant but it is included to give a general schema.
The program above creates a finite cyclic data structure.
If we wanted to expand out the cyclic structure, into an
infinite copy, f could be redefined as follows:

\begin{verbatim}
  f empty = empty ||
  f (node a subtree sibs) =
    let others = f sibs
    in node a (f subtree) others
\end{verbatim}

Solutions to the various problems discussed below are
formed only by redefining f in variations on the above.
Apart from the changes that this entails the schema is unaltered in
each case.

\subsection{Permutations}
\label{Perms}

A common method of generating permutations is to extend
partial permutations, beginning with the empty sequence.
If abcde is a partial permutation it can be extended with X to
abcdeX provided that X differs from a, b, c, d and e.
Equivalently, it can be extended provided that bcdeX is a
partial permutation and provided that X differs from a.
Note that bcde is already a partial permutation because abcde is.
If a tree is used to hold the permutations then bcdeX, being
shorter than abcdeX, must occur in the tree at the previous
level, if it is indeed a partial permutation.
If we read permutations as paths from the
root of the tree, and identify a node
with the path to it, the subtree of abcde is a pruned version of
the subtree of bcde with all occurrences of a filtered out or
banned.
We call bcde the shadow of abcde.
The shadow of bcde is cde and so on.
Coding these ideas into function f
of the schema in the previous section gives the following program.

\begin{verbatim}
  let build n =
    let rec
      T = toplevel 1

    and toplevel m =
      if m>n then empty
      else node m (f m T)
                  (toplevel (m+1))

    and f banned empty = empty ||
                      --f banned shadow
      f banned (node a subtree sibs) =
        let others = f banned sibs
        in if a=banned then others
                      --prune a's subtree
           else node a (f banned subtree)
                       others  --no pruning
    in T
\end{verbatim}

Function f has gained an extra parameter for the banned
element and performs a filtering operation on the shadow tree.
A single test, a=banned, tells if a permutation can be extended
with a given value, all other exclusions being implicit in the
tree structure.

As the permutation tree is traversed it is gradually evaluated.
If for example the first permutation, 123, from build 3
were printed, the evaluated portion would be:

\begin{verbatim}
                   .
                .  .  .
             .     .     .
          .        .        .
         1         2         3
        . .       . .       . .
       .   .     .   .     .   .
       2   ?     1   3     1   2
       .         .   .     .   .
       .         .   .     .   .
       .         ?   .     ?   ?
       3             1
\end{verbatim}

Note, `?' is used to denote an unevaluated subtree.
Recall that we read a permutation, such as 123,
as a path from the root
of the tree and identify a node with the path to it.
The shadow of 123 is 23.
The subtree of 23 is `node 1 ? empty' and the 1
is banned for 123, so 123 is a complete permutation.
The shadow of 23 is 3.
The subtree of 3 is `node 1 ? (node 2 ? empty)'
and the second branch is banned beneath 23 so its
subtree is `node 1 ? empty'.

The shadow of a path grows with the path, one step behind it.
The structure of the tree stores the results of many past tests
so that only a single extra test is performed to add a new node.
(This is not a big issue in permutation generation but there are
cases where it is.)
Note that Topor \cite{Top82} has examined the space
complexity of functional programs for generating
permutations represented as linear linked lists.

\subsection{N-Queens}
\label{NQ}

The well known n-queens problem is to place n queens on an
n$\times$n chess board so that no two queens threaten each other.
Each queen must be on a separate row, column and diagonal
and this property is an invariant that must be maintained as
partial solutions are extended.
The fastest imperative solutions \cite{Roh83} are based on
permutation generators.
A board is represented by the permutation of rows that the
queens on the columns occupy. This representation automatically
ensures the separate row and column parts of the invariant.
Here we observe that a partial solution abcde can be
extended to a partial solution abcdeX if and only if bcdeX is
also apartial solution and a and x are on separate diagonals and rows.
By using shadows, X need only be tested against a's
diagonals as the results of the other diagonal tests against other
queens are already encoded in the shadow tree and do not need
to be repeated.
Again, the required program is a modification
of the general schema with f redefined.
Function f gains a new
parameter col, being the current column number.

\begin{verbatim}
let build n =
  let rec
    T = toplevel

  and toplevel m =
    if m>n then empty
    else node m (f 1 m T) (toplevel(m+1))

  and f col banned empty = empty ||
    f col banned (node a subtree sibs) =
      let others = f col banned sibs
      in if member banned [a, a+col, a-col]
         then others  --prune
         else node a
                   (f(col+l) banned subtree)
                   others --no prune
  in T
\end{verbatim}

The standard function member tests the membership of an element in a list.
Note that the test member banned [a, a+col, a-col]
is an amalgam of the old permutation test, a=banned,
and the new diagonal test.

\subsection{Irreducible or Good Sequences}
\label{Irred}

Axel Thue \cite{Hed67} defined the notion of an irreducible
sequence in a series of papers in the period 1906-1914
Dijkstra \cite{Dij72} later called these `good sequences'
and used them in an exercise in structured programming.
A sequence over the alphabet [1, 2, 3], or in general over
[1,...,n], is irreducible if and only if it contains no adjacent
subsequences that are identical.
For example, 1213121 is irreducible but 12132131
is not because 213 is immediately repeated.

It is easy to see that (i) a sequence abcdeX of even length
is irreducible if the shorter sequences abcde and bcdeX are
irreducible and the two halves abc and deX are unequal and
(ii) a sequence abcdefX of odd length is irreducible if abcdef and
bcdefX are irreducible.
This enables a circular program to be
written for a tree representing all the irreducible sequences.
For example, the shadow of abcde (of odd length) is bcde.
Assuming that abcde is irreducible, abcdeX is irreducible if
and only if bcdeX is irreducible, if and only if X is
a descendant of bcde.

\begin{verbatim}
let build n =
  let rec T = toplevel 1

  and toplevel m =
      if m>n then empty
      else node m (f 2 [m] T)
                  (toplevel (m+1))

  and f len seq empty = empty ||
    f len seq (node a subtree sibs) =
      let others = f len seq sibs
      and seq2 = a.seq
      in if even len & repeated (len/2) seq2
         then others --prune
         else node a (f(len+l) seq2 subtree)
              others  --don't prune
  in T
\end{verbatim}

A new parameter seq carries the particular sequence forward
as f descends through the tree.
When the length len is
even, a test is made that the sequence a.seq is not the
concatenation of two sequences of length len/2.
Function repeated performs this test in $O(len)$ time and has an obvious
definition.
It is not necessary to test for any shorter repeats.
These are implicitly ruled out by the use of the shadow to
generate subtrees.
The test would be more complex without
this information.
That portion of the tree that is evaluated in
order to print the first irreducible sequence of length five is
shown below:

\begin{verbatim}
                       .
                   .   .   .
               .       .       .
           .           .           .
          1            2            3
        .   .         .            .
      .       .      .            .
      2       3      1            1
     .       .        .
    .       .          .
    1       1          3
     .                .
      .              .
      3              1
     .
    .
    1
\end{verbatim}

The shadow of 12131 is 2131 whose shadow is 131 whose
shadow is 31 whose shadow is~1.

\section{Conclusion}
\label{Conclusion}

Programmers are familiar with recursive functions but recursive
or self-referential data structures used in circular programs are rare.
Circular programs are very powerful enabling
many infinite structures to be efficiently defined.
They often remove the need for intermediate structures and for
repeated calculations.
They can be used safely provided that later
values depend only on earlier ones in the structure.
Memo-structures can be formed by a data structure and function
defined using mutual recursion.
Explicit circular search-trees
can reduce the number of tests performed in constraint-satisfaction
problems.
As usual, there is a trade-off of time against space.


\bibliographystyle{plainurl}

\bibliography{paper}

\section*{Biographical Note}

Lloyd Allison [was] a Reader at the Department of
Computer Science, Monash University. He received his BA from
Cambridge University in 1973 and was awarded his MSc
(distinction) in Computer Science from London University
Institute of Computer Science in 1974 and his PhD in Computer
Science from Manchester University in 1976.
His book
``A Practical Introduction to Denotational Semantics'' [14]
was published by Cambridge University Press in 1986.

\noindent
\rule{\linewidth}{1pt}

\section*{This copy}

The paper above originally appeared [15] in the
`Australian Computer Journal' (ISSN 0004-8917).
The journal was published by the Australian Computer Society
from 1967 to 1999, later as the
`Journal of Research and Practice in Information and Technology'
(ISSN 1443-458X)
-- see \myhref{https://catalogue.nla.gov.au/Record/563887}{NLA}(6/2022).
The typesetting of this copy differs from that in the ACJ
as the original LaTex(?) is no longer available
--~\myhref{https://www.cantab.net/users/mmlist/ll/}{L.A.}

\begin{itemize}
\item[{[}14{]}] L. Allison.
  \textit{A Practical Introduction to Denotational Semantics}.
  Cambridge University Press, 1986.
  \myhref{https://doi.org/10.1017/CBO9781139171892}{doi:10.1017/CBO9781139171892}
\item[{[}15{]}] L. Allison.
  Applications of recursively defined data structures.
  \textit{Australian Computer Journal}, 25(1):14--20,
  February 1993.
\end{itemize}

\end{document}